# Investigation of Early-Stage Breast Cancer Detection using Quantum Neural Network

Musaddiq Al Ali [a], Amjad Y. Sahib [b], Muazez Al Ali [c]

[a] Department of Advanced Science and Technology, Toyota Technological Institute, 2-12-1,Hisakata, Tenpaku-ku, Nagoya, Aichi 468-8511, Japan

[b] University of Wasit, College of engineering. Al Kut, Wasit, Iraq

[c] Al Ayen University, College of Dentistry, Nile St, Nasiriyah, Iraq

Corresponding author: alali@toyota-ti.ac.jp (Musaddiq Al Ali)

**Abstract**

Computer-aided image diagnostics (CAD) have been used in many fields of diagnostic medicine. It relies heavily on classical computer vision and artificial intelligence. Quantum neural network (QNN) has been introduced by many researchers around the world and presented recently by research corporations such as Microsoft, Google, and IBM. In this paper, the investigation of the validity of using the QNN algorithm for machine-based breast cancer detection was performed. To validate the learnability of the QNN, a series of learnability tests were performed alongside with classical convolutional neural network (CCNN). QNN is built using the Cirq library to perform the assimilation of quantum computation on classical computers. Series of investigations were performed to study the learnability characteristics of QNN and CCNN under the same computational conditions. The comparison was performed for real Mammogram data sets. The investigations showed success in terms of recognizing the data and training. Our work shows better performance of QNN in terms of successfully training and producing a valid model for smaller data set compared to CCNN.

## 1- Introduction

With the advancement in medical and engineering fields, novel solutions were implemented to facilitate patients' health care and prolong their life[1,2,11–20,3,21–26,4–10] . The use of computer use of computer-aided diagnostic (CAD) is an important topic in engineering-medical research[27,28]. Recently, many researchers investigated the concept of automating CAD by building self-learning algorithms based on machine learning [29][30][31]. In order to build a successful diagnostic model of medical images, classical machine learning is used. However, training classical machine learning is consuming a huge computational resource in terms of the data set preparation as well as computer resources for the training phase [31][32]. For the medical diagnostic field, data are mostly visual-based, such as X-rays, computed tomography scan and magnetic resonance imaging (MRI), etc. Therefore; computer vision tools are the most appropriate method to be used for CAD. Add to that, most data in modern diagnostic are computerized based [33]. As such, artificial intelligence is the most suitable to be used in the next generation of fully computerized CAD. Historically, CAD [34] has been approached in two sequential steps. The first step is to screen the data to detect the suspicious regions or what are so-called (Regions of interest) then, the region of interest will be "labeled" for the closest possible medical cases according to the corresponding disease likelihood. This is done by what so called expert systems [35][36]. Expert systems will assign diagnostic labels according to the probability of the region of interest diagnostic ascendingly, then eliminate medical cases of lower probabilities and then identify the highly likely disease. The problem with the use of expert systems is that they are built upon predetermined diagnostics for a fixed small number of cases and are mostly based on a wide variety of collective diagnostic data. For example, to diagnose a single case of apparent mass in appeared in mammography, a series of tests should be performed (Blood, tissue samples, multiple scans, etc.)

to be fed to the expert systems than gives the preliminary diagnosis. On the other hand, artificial intelligence (AI) has great potential to replace traditional expert systems and allows one to reach a preliminary diagnosis in a very short time. AI is presented itself as a powerful tool for image classification and medical identification due to its characteristics of transforming representing information sets of data (database) into structured matrices of simple units containing weighted partial differential equations (PDEs). Weights then will be tuned to draw learning pathways as an intuitive mimic of the learning process of the biological neural system (i.e., the brain). The satisfying amount of learning data (i.e., raining data) to provide a robust outcome or valid model is depending on the AI model design and the problem itself. For data selection, two major aspects should be taken into consideration; the first one is identifying the objective of data to be trained for. In other words, what are the systematic methodologies of the doctor to make diagnostics based on such data. For example, in microscopic cultures, counting cells is one of the diagnostic tools; therefore, counting will necessitate isolating (through image segmentation) as well as labeling the targeted cells in each picture before submitting the data to the deep learning model. The second aspect is the data set size and number of items for each label. As it has been expressed before, training robustness is susceptible to data set size and the balance of data distribution in each label. The bottleneck of creating a strong and highly trusted CAD is the computational power. For classical machine learning (which is performed on nowadays classical computers that use the binary system [0,1]), the researchers reach the computation limitations to modeling real-life problems such as biochemical interactions and immune system interaction with infection. Some algorithms are invented to rework the data feed and training process to optimize workflow for the available resources. Furthermore, some hardware-based methods are introduced to give the classical systems the needed dynamic storage to allocate the data feeds for the processing units as well as storing the tensor from the data (such as Intel Obtain-based modules and tensor processors). However, the largest computer (supercomputer) is still facing huge challenges to simulate the simulation tasks mentioned earlier. As such, in the medical field, classical computers are still facing serious limitations in terms of attaining good diagnoses compared to well-trained doctors with the same amount of data. To overcome classical computers' limitations, researchers are working on developing so-called quantum computers. Quantum computers are applying the principles of entanglement and superpositions to the data unit, i.e., the bit to build the quantum bit. There are several approaches to physically implementing quantum computers such as photonic and silicon-based circuitry. However, the development is not moving fast as it has been anticipated due to entropy and noise problems. On the other hand, the logical aspects of a quantum computer are mature enough to be implemented as soon as a full-scale quantum computer is available. The advantages of a quantum computer can be shown by the phenomenal calculation speed and the data amount to be handled in a single calculation. The high speed and huge data processing ability of quantum computers are due to the nature of the method of conveying information itself, such that, instead of the binary representation of the data (i.e., classical bit), quantum bits within the quantum gates exchange information timelessly. The property of a quantum state makes one quantum computer hold calculation power equivalent to all existing classical computers (i.e., quantum supremacy) [37].

Quantum computers and computation are superior due to the nature of quantum information and quantum logic. Although the quantum computation field of study is relatively a new branch of applied mathematics, however; the physical and mathematical advancements of quantum computers are motivating the researchers to race toward implementing a unique type of logic gates and physical hardware. Nowadays, limited quantum bits computers are already serving in specific high technology applications such as pharmacy, and chemical interactions in next-generation batteries. It is anticipated that quantum computers will be available for public use with full capability at the end of the 21$^{st}$ century. The use of quantum computers for medical diagnostics can play a vital role in terms of the implementation of fast and systematic machine-based diagnostics of malignant tumors which are treatable if they can be diagnosed in the early stages. Breast cancer is the most frequent malignant tumor amongst women. It is a dominant cause of female mortality and is considered a serious public health problem all over the world. Current treatments for breast cancer include surgery, chemotherapy, immunotherapy, and radiation therapy. Breast cancer incidence and death rates increase with age but are mortality rate decrease significantly if it can be detected in the early stages, before the metastasizing phase. The eradication and therapeutic success of breast cancer are related to tumor stratification and dissemination. Breast tumors whether they are benign or malignant are distinguished into four major classes, based on size, age, node involvement, and tumor grade. These stages are 1; consists of the well-defined and localized tumor mass, characterized by poor invasion properties. Stage 2 and 3, corresponds to an increased tumor volume and acquisition of invasive phenotype. The metastasis dissemination and huge tumor size with invasive phenotype are classified as stage 4. Chemotherapy, radiation, and targeted therapies have made major advances in patient management over the past decades, but refractory diseases and recurrence remain common. The early-stage diagnostic will lead to treating cancer before the metastasis stage, at which cancer will attack different organs by migrating through lymph nodes. A mammogram is one of the popular tests for breast cancer early detection. Mammograms have been used efficiently to reduce the mortality rate of women with breast cancer. Early detection is based on the oncologist's exam of the x-ray image and then examining the suspicious tissue by taking a biopsy. However, due to the limitation of highly trained medical staff, cancer false positive is common in mammogram image detection as well as false negative [38–40] due to the misconception of less-skilled eyes for the masses in the image. Another point worth mentioning is that mammogram is an X-ray image that take short time to be made, yet the speed of mass examination is related to how many doctors exist and their level of experience. Therefore, CAD is a promotion point as a necessary way to reduce diagnostic time [41–43] and decrease cost per test by reducing the number of specialists in the mammogram testing unit in hospitals (This reduction will lead to remobilizing the manpower to other sections within the hospital which will lead to double the efficiency of the hospital's workforce) as well as the reduction of the number of false-positive by double-checking the X-ray image by the doctor and the computer.

This paper is examining the use of QNN to build an "ultra-intelligent machine". To attain this goal, the following Key questions should be answered: Firstly, how to build a QNN and how to transform the classical data from plane mammogram images to become quantum data. The second key question is how to evaluate the learnability of QNN. The third question is, how is implementing a benchmarking of learnability evaluation of QNN? The final question is: can we deduce that, QNN is able to perform successful mammography diagnostic in the future according to the current investigation?

As such, this study introduced the learnability factors. Furthermore, a series of numerical investigations were conducted to examine the various aspects that govern AI with QNN. Moreover, a hybrid model of CCNN and QNN is introduced to allow the implementation of quantum logic in the near future instead of waiting for a fully capable quantum computer to conduct breast cancer early detection for mammography.

the layout of the paper is the following. Section 2 is discussing breast cancer and diagnostic principles. whereas Section 3 deals with the learning of classical and quantum neural networks. Section 4 focuses on quantum neural network learning. Section 5 discusses the layout of quantum and classical neural networks and mammogram data structure. In section 6, we will present the results and discuss the outcome. Finally, the conclusions are presented in section 7.

## 2- Brief of breast cancer biology and transcriptional regulation

Breasts are made up of connective, glandular, and fatty tissues that have lobes, lobules, ducts, areola, and nipple[44,45]. These organs consist of a uniform structure of epithelial cells that secrete and produce milk after childbirth. Whenever there is a morphologic or functional alteration within its uniform epithelial structures, tumor initiation develops and later forms a mass of multiple populations of cells capable of evading physiological cell death. The changes in gene expression patterns seen in breast cancer[46] have provided evidence of epigenetic, genetic, or post-translational altered expression of certain proteins, like transcription factors, co-regulators, and histone enzymes that order DNA into structural units according to a recent study. These proteins play a crucial role in the expression of genes that results in the susceptibility of a healthy cell to the transformation of a malignant cell. Among the first altered transcriptional regulation found in breast cancer were the overexpression and gene amplification of estrogen receptor alpha (ERα) and avian myelocytomatosis viral oncogene homologue factor (c-myc). These two oncoproteins were found to be associated with abnormal cell division and replication within the breast. Additional studies have identified inherited/acquired altered gene expression as a detectable cause of carcinogenesis of breast tissue. This arises after a study of some essential genes involved in cellular processes and maintenance were found to be mutated at the germ cell level. Next-generation sequencing analysis also found higher penetrance mutations in breast cancer 1 (BRCA1), tumor protein p53, mitogen-activated protein kinase 1 (MAP3K1), retinoblastoma 1 (RB1), phosphatidylinositol-4, 5-bisphosphate 3-kinase catalytic subunit alpha (PIK3CA), and GATA binding protein 3 (GATA-3) genes that result in breast cancer formation. Breast cancer diagnostics has two-stage. The first stage is identifying whether there is irregularity within breast tissues (Masses). The second stage is doing a biopsy of the suspicious mass. Abnormal masses within breast tissue are the significant features that determine the likelihood of cancer's existence in the breast. The first identification is shaped. The shape can be rounded or irregular. An irregular shape increases the likelihood of malignancy (Cancer). Margins of the masses are also important. The margin can be circumscribed, microlobulated, obscured (partially hidden by adjacent tissue), indistinct (ill-defined), or spiculated. The likelihood of malignancy of circumscribed margins is low. The density of the mass is also a factor in early diagnosis. For example, if the mass is cystic or fibroadenomas mass. Other malignancy signs are neodensity, architectural distortion, or asymmetric density.

### 3- Classical and quantum neural network

Machine learning is the transcending form of the neural network, which was immersed first in the form of a "logic theorist program" that was invented in 1956 [47]. The neural network process starts by discretizing the problem to what so called neurons (X in Fig. 1). Multiple neurons (thousands and even millions) then will be stacked in a certain design to make the machine learning architecture. The information to be processed will be distributed on the neurons within the machine learning architecture, then fitting the response of each neuron to arbitrary function ψ. Functions output will be submitted to a comparator Ω which judges the data based on its probability function the gives the generalized output of the subsystems $\Psi$.

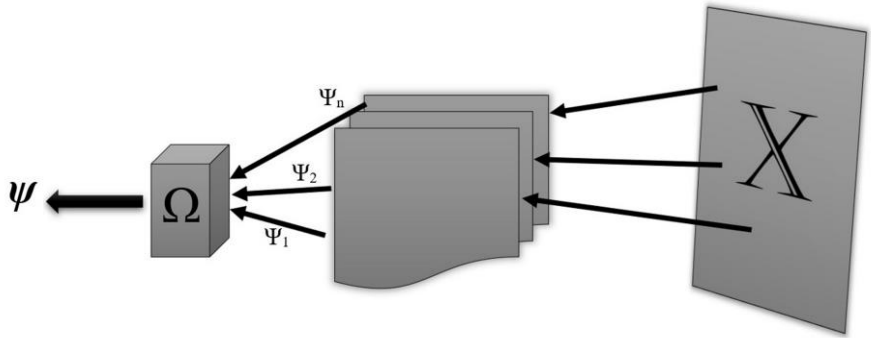

**Fig. 1. Machine intelligence process unit (neuron).**

In summary, machine learning is a network of node clusters; each node has a weighted subfunction to be tuned. This tuneable subfunction is representative of a fraction of the resulting model. The node's output mimics neural cells by adopting an activation function to control the output with predefined criteria. Activation function can take the form of rectified linear, Sigmoid, or hyperbolic. The stacks are consisting of an aggregation of neurons in layers. The layers may take many formations according to the machine learning architecture (e.g., dense and convolution layers). Training of the neural network is performed by tuning each neuron's weights for the optimum value that fits the training data. Here, optimization algorithm selection plays a significant role in training success. With increasing the training rate, data loss for each prediction will drop. If the data is insufficient or the neural network is designed poorly, the fit convergence cannot reach sufficient value. In this case, the underfitting problem is occurring. Contrary to the underfitting problem, the overfitting problem occurs when the convergence is satisfied for the model; however, the neural network has the poor capability to predict and recognize unique and new input data that is not in the training set.

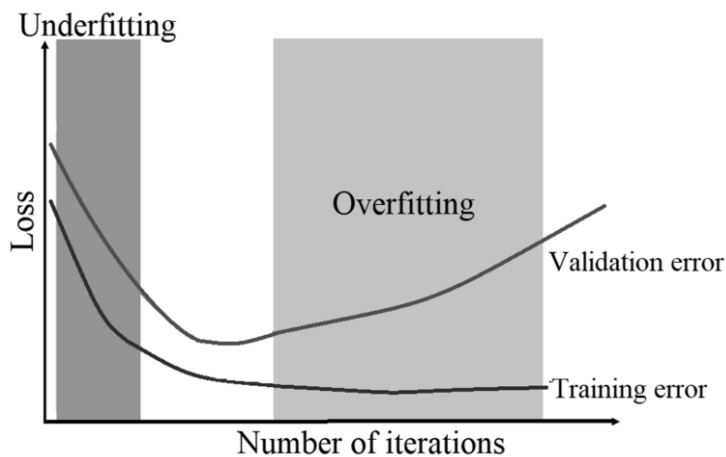

**Fig. 2. Under and overfitting of neural network.**

Quantum machine learning is a subject of the quantum computation branch, that has been gaining attention in recent decades, which emerged from quantum computations. Quantum computation has emerged from the statistical implementation of the work of quantum theory [48][49]. Moreover, the concept of a quantum computer was researched by many researchers, such as Benioff [50] and Feynman [51,52]. Quantum computers, due to their design, have a unique ability to perform extensive data analysis that is beyond classical computers' capability [53]. Quantum computers unit information, i.e., a quantum bit (Qbit), shares the same characteristics of quantum concepts, such that it is in a simple state of a linear superposition of (0) and (1) [54,55]. As such, the simple quantum state can be represented in a Dirac notation as:

$$\text{simplistic quantum state} = \begin{cases} |0\rangle \begin{bmatrix} 1 \\ 0 \end{bmatrix} \\ |1\rangle \begin{bmatrix} 0 \\ 1 \end{bmatrix} \end{cases} \quad (1)$$

This means that a linear combination of $|0\rangle$ and $|1\rangle$ can adequately describe in Qbit as:

$$\alpha|0\rangle + \beta|1\rangle \quad (2)$$

Where $\alpha$ and $\beta$ are the vector amplitudes representation of the basic units (i.e., the 0 and 1) in Hilbert space, such that ($\alpha^2 + \beta^2 = 1$). This vector is a representation of two original classical states i.e., 0 and 1 of phase $\phi$. Such a sphere of unity will give the qubit the ability to represent a probability of the included basic units i.e., quantum parallelism. To get the value of the quantum state of a qubit, it is necessary to do what so-called collapse. Collapse is an irreversible process in which the value of the qubit will be measured. Here, the output of quantum computation will be presented in a similar analogy to a classical computer.

To perform quantum computation within the quantum computer, a special logic operator controller i.e., quantum gates is needed to interpret and operate a program using the previously prescribed Qbit. Quantum gates are represented mathematically with a set of matrices representing the probability of a unified state within the state. For example, quantum NOT gate (as shown in equation 3) and the Hadamard gate (as shown in equation 4)

$$X - \boxed{X} - \begin{bmatrix} 0 & 1 \\ 1 & 0 \end{bmatrix} \quad (3)$$

$$H|0\rangle = \frac{1}{\sqrt{2}} \begin{bmatrix} 1 & 1 \\ 1 & -1 \end{bmatrix} \begin{bmatrix} 1 \\ 0 \end{bmatrix} = \frac{1}{\sqrt{2}} \begin{bmatrix} 1 \\ 1 \end{bmatrix} = \frac{1}{\sqrt{2}}|0\rangle + \frac{1}{\sqrt{2}}|1\rangle \quad (4)$$

4- **Machine learning and quantum computers**

The idea of computer intelligence was born with the emergence of modern computers in the mid of 20th century. The early modeling of machine learning has emerged as the perceptron algorithm [56]. Many tried to solve the idea of the perceptron learning model that shaped the modern view of machine learning [57][58][59][60]. The modern concept of machine learning is based on transforming the data sets into matrix formation of sub equations to be shaped by heuristic [6][7] or metaheuristic [61] optimized fitting. Then after training the resulting model will be evaluated against a separate controlled set of data (usually new data that the model did not train for before). Neural networks, whether they are classical or quantum, are built from the aggregates of layers that consist of numerous neurons. The layers consist of building blocks containing weighted mathematical fitting criteria. The weights and the bias of the fitting criteria are calculated using successive training and adjusting the optimization process to have the optimal value of the constants that are shaping the final values of the regulation factor of the neurons (i.e., weights and biases). In the case of CCNN, the successive layers are utilized to obtain better feature extraction, then delivered to a flatted output layers to connect all the neurons in a way that allows the computer to identify the data correctly. In the case of QNN, quantizing these building blocks is necessary to perform quantum machine learning. Quantum machine learning (QML) is based on quantum computation implementation with the same intuition of classical machine learning but with the supremacy of quantum search algorithms. The quantum neural network (QNN) is a powerful tool in modern computer vision-based AI [62]. From its name, it used quantum-based neurons to build a series of layers based on quantum gates. These layers will be used to extract the image's features whether it is a non-constrained training (such that the neural network has no feature labels within the single image to guide specific features of the training, as in the present case in this paper) or a constraint training which needs labeling each needed feature with each image. In general, the learnability of a neural network can be tested initially by monitoring the fitting process (Fig. 2.). Moreover, it is important to mention that, in our research and for image recognition using classical machine learning, the convolutional neural network technique is implemented successfully to compress the image while maintaining the learnable data (i.e., the features). However, prior compression of the image is necessary for managing computer resources and maintaining the calculation without RAM overflow.

5- **Materials and methods**

Mammogram image data set is provided by [63] for 6 breast abnormality classes: calcification, circumscribed masses, spiculate masses, architectural distortion, asymmetry, and healthy breast images. The data set consists of the training set of 45000 images for all the previously mentioned 6 classes and 7500 images as the validation set. Each image is of size 1080 by 1080 pixels. MATLAB was implemented to investigate CNN design and perform training. Because quantum-based libraries provided by quantum computer service providers (such as IBM and GOOGLE) are built with Python, the Python version of the optimal MATLAB CNN was used as a baseline comparison program with QNN. Quantum gates formation and QNN were built using python language. For classical coding, CNN was built with multi-stages of convolutional layers to compress the data stream to the throttle point (as shown in the machine learning architecture in Fig. 3). Images are needed to be resized to decrease the pixel numbers to smooth the training; add to that, reducing pixel numbers is necessary to perform training with reasonable computer resources. Using raw images of high resolution will necessitate an increase in the node's number inside the neural network, which will not necessarily improve the model recognition ability.

As such, in our research, input images were compressed to 256 by 256 pixels and uploaded using the computer unified device architecture's library (CUDA) to enable the parallelization of tensor multiplication. Several CNN models were built to adjust the optimal neuron number, starting from 50 million neurons to 2 million neurons.

The best CNN design and the optimal number of neurons are evaluated based on the model's learnability and validation. Learnability is the most important aspect of machine learning, such that it will show if the program or the set is in fact, machine intelligence or not. The learnability can be evaluated by the behavior of the machine learning to data and the change of its architecture in a way that it should always show the symptoms of acquiring the information. One of the easiest symptoms to be measured is the under and overfitting of the machine model to the training data. For example, a neural network by design might not converge sufficiently to give a valid model. This phenomenon can be referred to as underfitting. Underfitting can happen for many reasons, such as poor design models or insufficient data. On the other hand, if the learning rate is converged too much, the model recognizes the training set flawlessly. Yet, it lacks the ability to recognize the new images that the model is never experienced in the training phase. This phenomenon is called overfitting. Overfitting can happen due to the lack of trainable neurons in each training cycle. The transformation from underfitting to overfitting of the neural network strongly indicates the successful learnability of the implemented network design.

On the other hand, our QNN was built using the Python language framework associated with the Cirq library as a quantum circuit framework [44][45][66]. Cirq library allows to perform of quantum computation on classical computers by assimilating the Qbit by its representative classical bits. This is the only way to test full-scale quantum programming due to the lack of usable quantum computers at the current time. However, due to the nature of Qbit, a huge amount of storage is needed due to the complex nature of Qbit and its phenomenal information capacity. As such, performing full-scale, and high or even moderate resolution image training with QNN on classical computers is extremely difficult due to the lack of computational resources to perform superposition, therefore; images must be sufficiently compressed to allow the computer to perform a quantum-based search algorithm.

In this paper, QNN work frame is as follows: The first step is to load the images into the model and store the data in the RAM as tensors of classical bits. Because quantum computation handles only quantum data, it is inevitable to transform classical bits into Qbit using the Cirq library, which is the second step of the QNN work-frame. Transformation is performed by mapping bits into a Qbit matrix using the transform function based on tensor products of Bits [67]. The third step is to build the quantum circuit of sequential quantum gate formations, as shown in Fig. 4. In this paper, a quantum circuit consists of a series of Ising gates that show good recognition ability. Starting from 50 x 50 pixels image compression, QNN simulation consumed computer resources before training started. In this research for our QNN design investigation, our model could not work for the full training and validation sets unless the image compression was extremely to 4x4 pixels (as presented in Fig. 5). After compression, the classical bits are transformed to the Qbit using the previously mentioned method.

After wrapping up the compressed training data set in terms of the quantum state (i.e., Qbit), the training of the qubit will start. The final training matrix will be exported and saved to be implemented in the prediction program. Our QNN algorithm is summarized in Fig. 6.

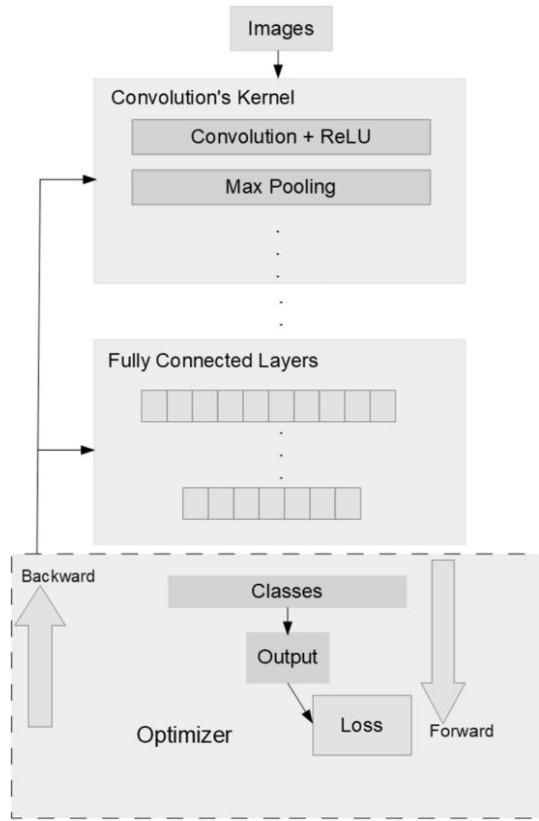

**Fig. 3. Classical convolution neural network architecture.**

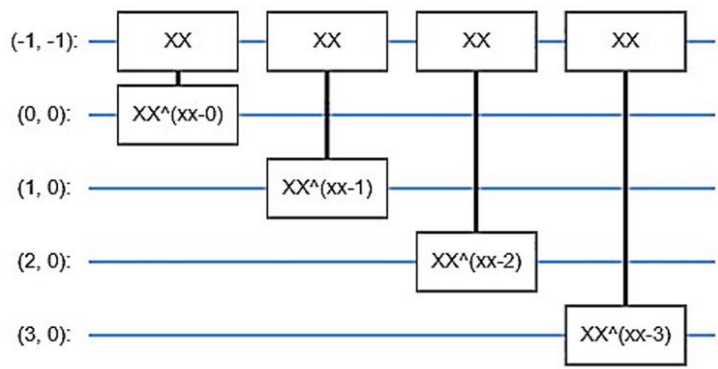

**Fig. 4. Ising quantum computation circuit for mammogram image-based cancer detection.**

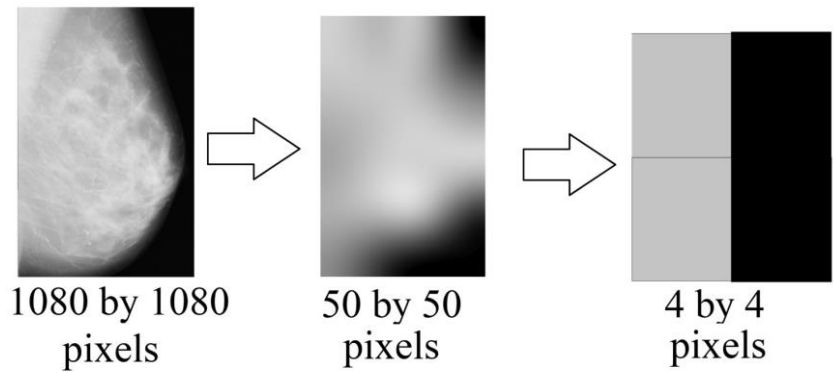

**Fig. 5. Image compression representation before full-scale QNN training.**

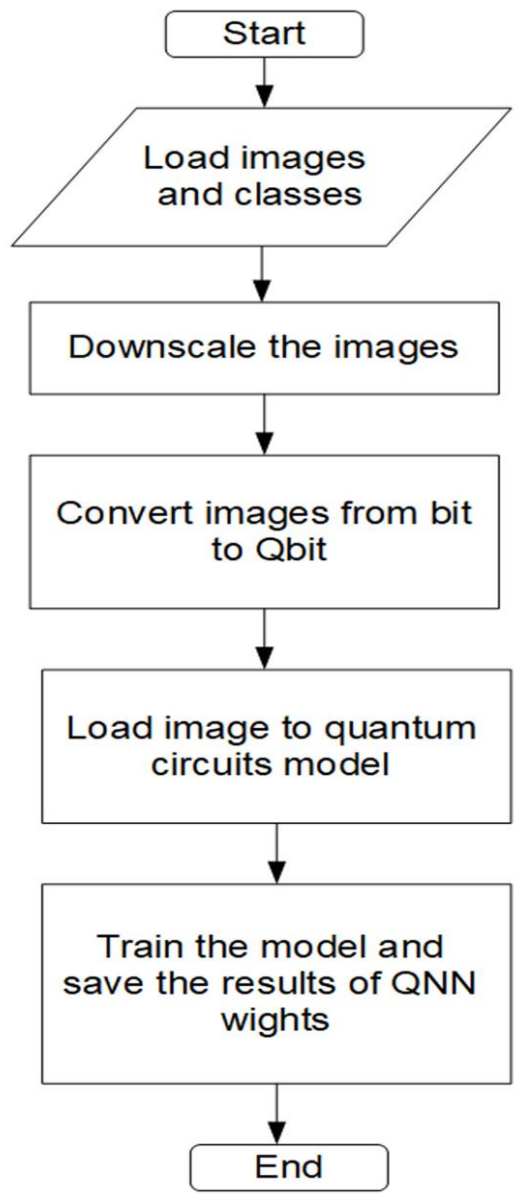

**Fig. 6. QNN identification algorithm.**

## 6- Results and discussion

In this section we are showing the results of our investigation of the use of CCNN and QNN

**6-1- The investigation of CCNN models**

As been previously explained in the material and method section, CCNN was examined through several designs by increasing the neuron count from 2 million to 50 million. A high number of neurons is needed, so mammogram image classification is difficult even for human eyes. On 50 million neurons, the model show overfitting, as shown in Fig. 7.

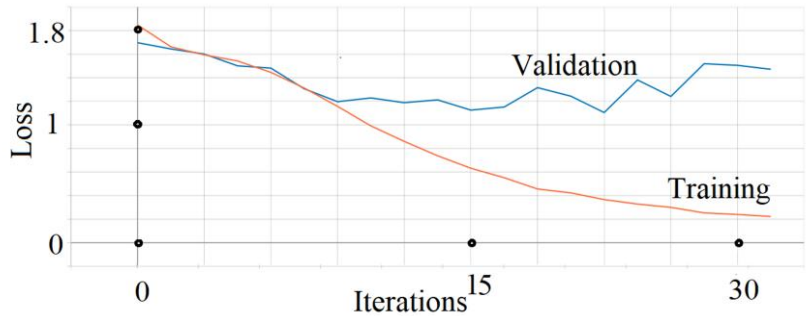

**Fig. 7. Classical neural network overfitting.**

While with 2 million neurons, the model shows underfitting behaviour (as shown in Fig. 8).

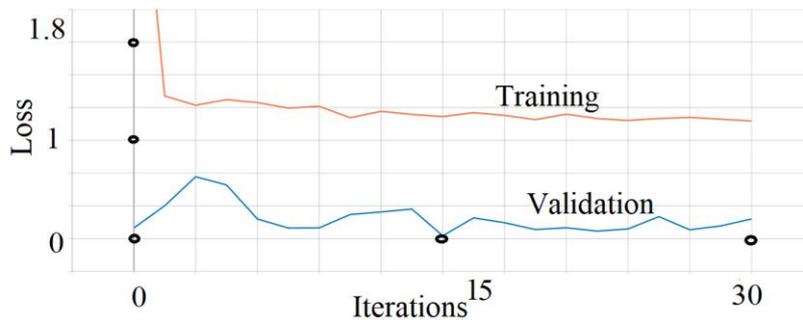

**Fig. 8. Classical neural network underfitting.**

Such a shift from underfitting to overfitting proves that the CCNN has a successfully learnability [68]. As shown in Fig. 9, the best natural network was the 1.4 million neurons model, while 50 million nodes now showed more than 53% accuracy (Fig. 10).

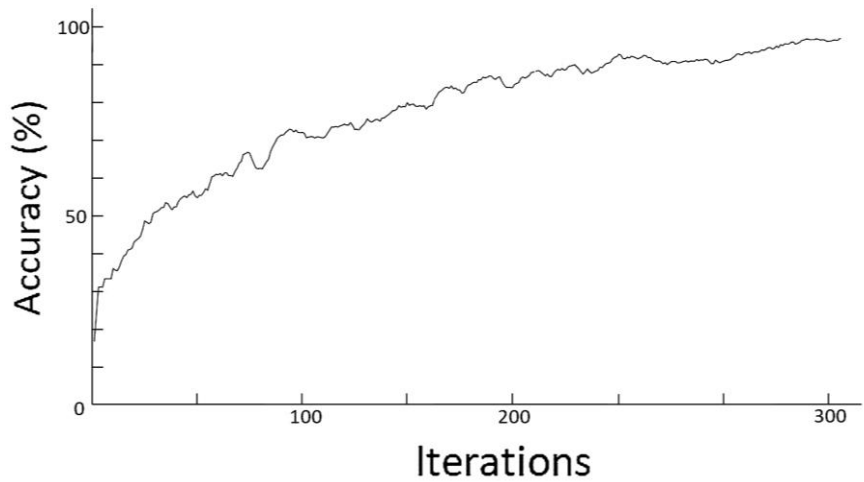

**Fig. 9 Accuracy history of the best classical neural network model.**

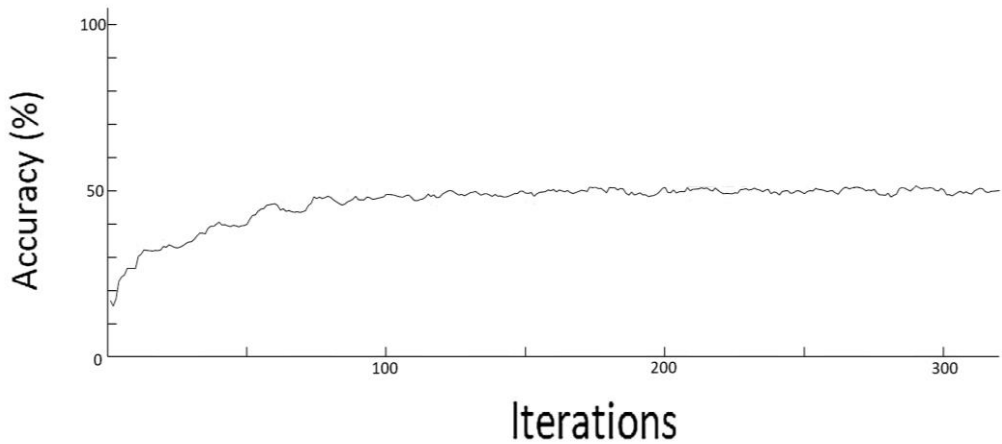

**Fig. 10 Accuracy history of the under-fitted classical neural network model.**

**6-2- The investigation of QNN models**

The capacity to learn is the most crucial component that measures the feasibility of using QNN as a mammogram diagnostic tool. The machine learning's response to input and the modification of its design in such a manner that it consistently displays the signs of learning may be used to assess the learnability. The under- and overfitting of the machine model to the training data are used as intuitive signs to validate the QNN learnability. As such, using the same analogy to examine QNN as an image classifier, a model showed a transformation from underfitting (Fig. 11) to overfitting (Fig. 12).

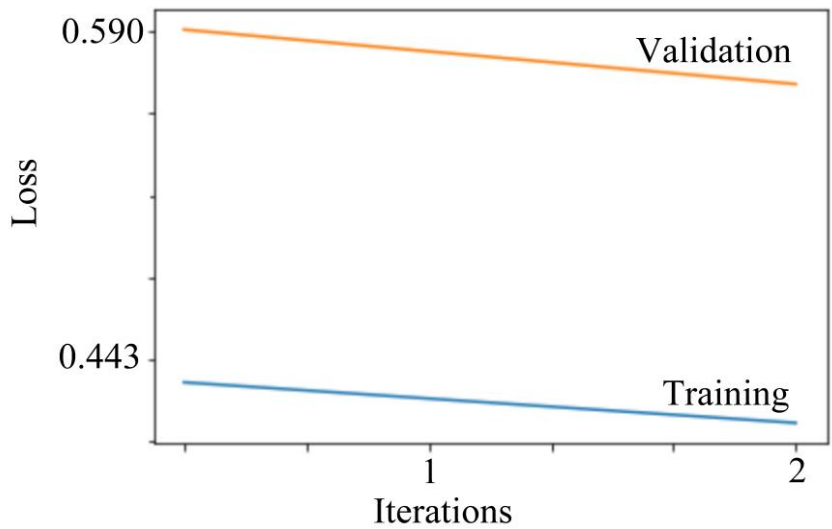

**Fig. 11. Quantum neural network underfitting.**

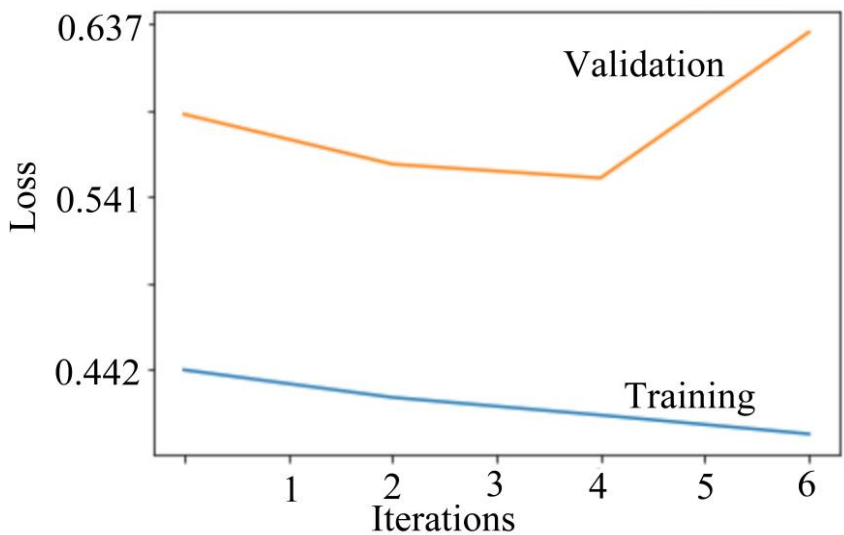

**Fig. 12. Quantum neural network overfitting.**

As mentioned earlier, because QNN was simulated on a classical computer, the image must be highly compressed to give the required RAM and CPU resources for constructing a Qbit. Although the mammogram images were highly compressed, QNN could reach converged after 5 training trials to achieve 58% accuracy (Fig. 13) with stead learning speed. Even though the training of highly compressed images, QNN showed sign of learnability by transforming the loss from underfitting to overfitting. This sign has proven the ability to use a large model of the QNN on a full-scale quantum computer to produce a robust CAD model. Also, QNN showed a high tendency to achieve a high fitting level with just several training cycles. This is an anticipated result due to the nature of the quantum gate operations.

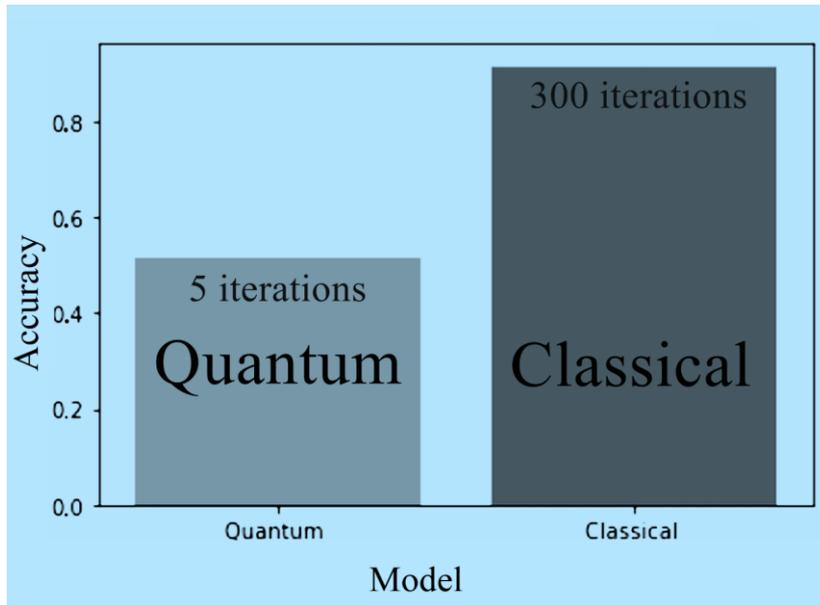

**Fig. 13. Classical and quantum network predictability.**

**6-3- The investigation of CCNN and QNN models using same data set**

To verify the learning ability of quantum and classical models in fair terms (i.e., examine both models under similar conditions), in this section we performed the CCNN and QNN evaluation for the training and validation data sets that have similar data compression and for the same number of samples for both models. Because it is impossible to train both QNN and CNN using the 45000 images with reasonable image compression, down-sampling of the full-scale data was examined to show which model is faster in achieving stable accuracy growth.

Two data sets were chosen for this study. The first set has 600 training images, and the second set of 1200 images. Both training cases have a validation set of 300 images. This choice was selected to ensure CCNN's fair learning chance and QNN stability during testing, so if the number of samples is chosen over 2000, it will lead to the overload of the computer memory. As such, for the data set of 600 images for the training set and 300 images for the validation test are compressed to 50 x 50 pixels (as shown in Fig. 14), and used for both CCNN and QNN models.

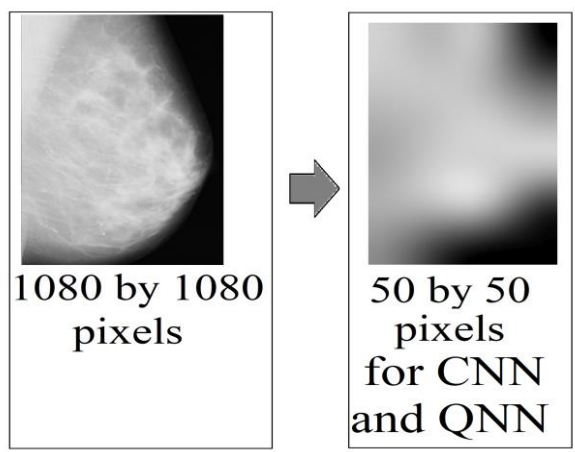

**Fig. 14. Image compression for the new data set.**

Training for this data set showed a remarkable convergence time of QNN compared to CCNN (as shown in Fig. 15). Validation accuracy for QNN showed superior performance compared with CCNN. CCNN showed fluctuation in validation accuracy, which can be interpreted as underfitting for the classical model.

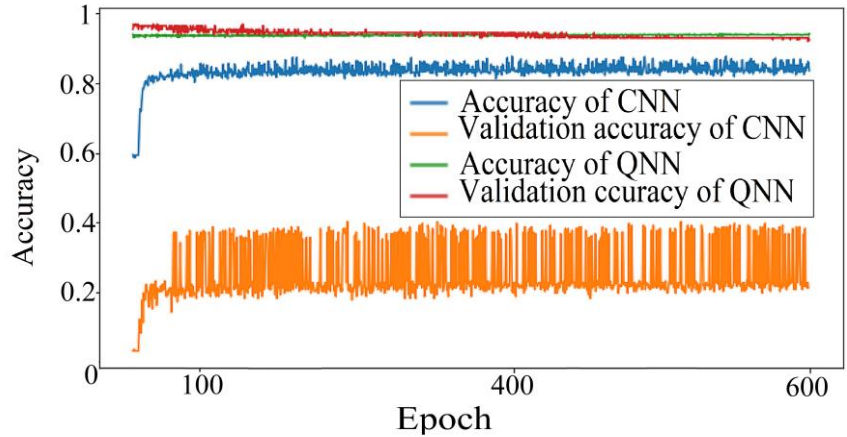

**Fig. 15. Classical and quantum network accuracy for 600 images data set.**

By increasing the training set to 1000 images to decrease the classical model's underfitting; QNN maintains its superiority compared to CCNN, as shown in Fig. 16. Mathematically speaking, the ability to parallelize the calculation process within the Qbit will lead to decreasing the nonlinearity due to continuous bidirectional feedback of neurons, as well as compressing the calculation process with maintaining robustness. For the foregoing reasons, QNN can achieve rapid, stable accuracy with a shorter time and fewer samples than CCNN.

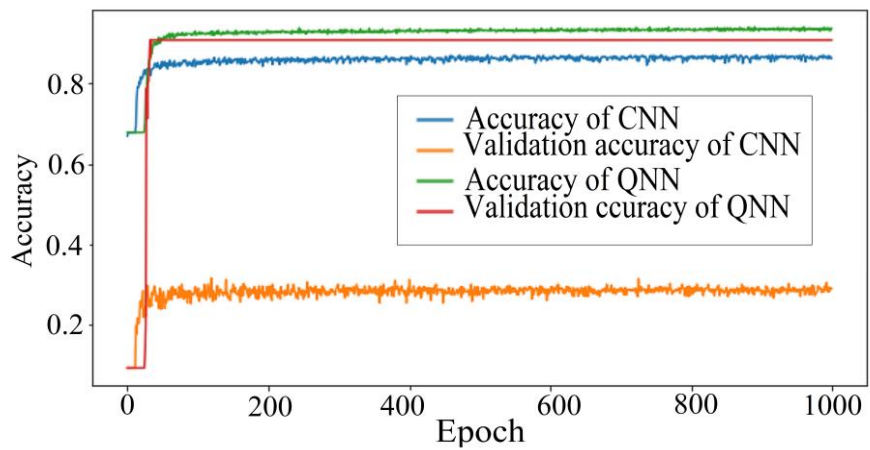

**Fig. 16. Classical and quantum network accuracy for 1000 images data set.**

It is valid to say that QNN can give a reasonable prediction model for a much smaller data set than CNN. Applying QNN on a fully capable quantum computer will provide faster training with smaller data set. Under such circumstances, QNN will provide fast mass medical imaging diagnostic with a minimum false negative.

## 7- Conclusion

In this research, we performed an investigation of early breast cancer detection using quantum computation by training mammogram images to a designed QNN model. The QNN model was successfully used to train actual mammogram data sets for comparison. The investigation demonstrated effectiveness in terms of training and data recognition. In comparison to CCNN as a mammogram image diagnostic, our study demonstrates the advantage of QNN in effectively training and delivering a viable model for less time and high accuracy. From this investigation, we conclude that the anticipated availability of real and robust quantum computers for commercial use will be a strong asset for robust mass mammogram detection successfully for a significantly short time with high accuracy. In other words, the main question discussed in this paper is whether it is possible to implement in theory, specific real data of early cancer detection in a quantum-based has been answered positively. It can be said that this objective was achieved by satisfying the criteria of learning for the designed QNN model. In addition, we conclude that QNN may provide a credible prediction model with a significantly smaller data set, unlike CCNN, which makes QNN an anticipated tool for performing mass medical imaging diagnostics with a low false-negative rate.

**Declaration of competing interest**

The authors declare that they have no known competing financial interests or personal relationships that could have appeared to influence the work reported in this paper.

**Data availability**

All results in this paper are calculated by using in-house MATLAB and Python codes. The code cannot be shared this time as it is used in ongoing work. However, all results can be reproduced by adopting the same assumptions.

**References**

[1] M. Fujioka, M. Shimoda, M. Al Ali, Concurrent shape optimization of a multiscale structure for controlling macrostructural stiffness, Struct. Multidiscip. Optim. 65 (2022) 1–27. https://doi.org/10.1007/s00158-022-03304-y.

[2] M. Al Ali, M. Shimoda, Toward Concurrent Multiscale Topology Optimization for High Heat Conductive and Light Weight Structure, in: S. Koshizuka (Ed.), 15th World Congr. Comput. Mech. 8th Asian Pacific Congr. Comput. Mech., CIMNE, 2022: p. 12. https://doi.org/10.23967/wccm-apcom.2022.118.

[3] M. Al Ali, M. Shimoda, Investigation of concurrent multiscale topology optimization for designing lightweight macrostructure with high thermal conductivity, Int. J. Therm. Sci. 179 (2022) 107653. https://doi.org/10.1016/j.ijthermalsci.2022.107653.

[4] N. Amoura, B. Benaissa, M. Al Ali, S. Khatir, Deep Neural Network and YUKI Algorithm for Inner Damage Characterization Based on Elastic Boundary Displacement, in: Proc. Int. Conf. Steel Compos. Eng. Struct. ICSCES 2022, 2023: pp. 220–233.

[5] M. Shimoda, M. Umemura, M. Al Ali, R. Tsukihara, Shape and topology optimization method for fiber placement design of CFRP plate and shell structures, Compos. Struct. 309 (2023) 116729. https://doi.org/10.1016/j.compstruct.2023.116729.

[6] R.S. Abass, M. Al Ali, M. Al Ali, Shape And Topology Optimization Design For Total Hip Joint Implant, in: World Congr. Eng. 2019, 2019.

[7] M. Fujioka, M. Shimoda, M. Al Ali, Concurrent Shape Optimization for Multiscale Structure with Desired Static Deformation, Proc. Comput. Mech. Conf. 2021.34 (2021) 3. https://doi.org/10.1299/jsmecmd.2021.34.003 (in Japanese).

[8] Musaddiq Al Ali, Toward fully autonomous structure design based on topology optimization and image processing, in: Proc. 6th IIAE Int. Conf. Intell. Syst. Image Process. 2018, 2018: pp. 1–7.

[9] M.S. Musaddiq Al Ali, Concurrent Multiscale Topology Optimization for Designing Displacement Inverter, in: 15th World Congr. Comput. Mech. 8th Asian Pacific Congr. Comput. Mech., 2022: pp. 1–10. https://doi.org/10.23967/wccm-apcom.2022.027.

[10] M. Fujioka, M. Shimoda, M. Al Ali, Shape optimization of periodic-microstructures for stiffness maximization of a macrostructure, Compos. Struct. 268 (2021) 113873. https://doi.org/10.1016/j.compstruct.2021.113873.

[11] P.M. Shimoda, M. Al Ali, C.A. Secondary, C. Author, M. Al Ali, M. Shimoda, M. Al Ali, Structural and Multidisciplinary Optimization Concurrent multiscale multiphysics topology optimization for porous composite structures under hygral loading, (n.d.).

[12] M.S. Musaddiq Al Ali, Toward Concurrent Multiscale Topology Optimization for High Heat Conductive and Light Weight Structure, in: 15th World Congr. Comput. Mech. 8th Asian Pacific Congr. Comput. Mech., 2022: pp. 1–12. https://doi.org/10.23967/wccm-apcom.2022.118.

[13]    M. Al Ali, Design offshore spherical tank support using shape optimization, in: Proc. 6th IIAE Int. Conf. Intell. Syst. Image Process., 2018.

[14]    M. Al Ali, M. Shimoda, B. Benaissa, M. Kobayashi, Concurrent Multiscale Hybrid Topology Optimization for Light Weight Porous Soft Robotic Hand with High Cellular Stiffness, in: Proc. Int. Conf. Steel Compos. Eng. Struct. ICSCES 2022, 2023: pp. 265–278. https://doi.org/10.1007/978-3-031-24041-6_22.

[15]    M. Al Ali, A. Takezawa, M. Kitamura, Comparative study of stress minimization using topology optimization and morphing based shape optimization comparative study of stress minimization using topology optimization and morphing based shape optimization, (2019).

[16]    M. Al Ali, M. Al Ali, A.Y. Sahib, R.S. Abbas, Design Micro-piezoelectric Actuated Gripper for Medical Applications, in: Proc. 6th IIAE Int. Conf. Ind. Appl. Eng. 2018, The Institute of Industrial Application Engineers, 2018: pp. 175–180. https://doi.org/10.12792/iciae2018.036.

[17]    M. Al Ali, M. Al Ali, R.S. Saleh, A.Y. Sahib, Fatigue Life Extending For Temporomandibular Plate Using Non Parametric Cascade Optimization, in: Proc. World Congr. Eng. 2019, 2019: pp. 547–553.

[18]    M. Al Ali, A. Takezawa, M. Kitamura, Comparative study of stress minimization using topology optimization and morphing based shape optimization comparative study of stress minimization using topology optimization and morphing based shape optimization, in: Asian Congr. Struct. Multidiscip. Optim., 2018.

[19]    R.S. Abass, M. Al Ali, M. Al Ali, Shape And Topology Optimization Design For Total Hip Joint Implant, in: Proc. World Congr. Eng., 2019.

[20]    M. Al Ali, A.Y. Sahib, M. Al Ali, Teeth implant design using weighted sum multi-objective function for topology optimization and real coding genetic algorithm, in: 6th IIAE Int. Conf. Ind. Appl. Eng. 2018, The Institute of Industrial Applications Engineers, Japan, 2018: pp. 182–188. https://doi.org/10.12792/iciae2018.037.

[21]    藤岡みなみ, 下田昌利, A.L.I. Musaddiq Al, 所望変形を実現するマルチスケール構造の同時形状最適化, 計算力学講演会講演論文集. 2021.34 (2021) 3. https://doi.org/10.1299/jsmecmd.2021.34.003.

[22]    M. Al Ali, M. Shimoda, Toward multiphysics multiscale concurrent topology optimization for lightweight structures with high heat conductivity and high stiffness using MATLAB, Struct. Multidiscip. Optim. 65 (2022) 1–26. https://doi.org/10.1007/s00158-022-03291-0.

[23]    M. Torisaki, M. Shimoda, M. Al Ali, Shape optimization method for strength design problem of microstructures in a multiscale structure, Int. J. Numer. Methods Eng. 124 (2023) 1748–1772. https://doi.org/10.1002/nme.7186.

[24]    M.A. Al-Ali, M.A. Al-Ali, A. Takezawa, M. Kitamura, Topology optimization and fatigue analysis of temporomandibular joint prosthesis, World J. Mech. 7 (2017) 323–339.

[25]    M. Al Ali, A.Y. Sahib, M. Al Ali, Design Light Weight Emergency Cot With Enhanced Spinal Immobilization Capability, in: 6th Asian/Australian Rotorcr. Forum Heli Japan, 2017: pp. 1–11.

[26]    M. Al Ali, Toward fully autonomous structure design based on topology optimization and image processing, in: Proc. 6th IIAE Int. Conf. Intell. Syst. Image Process., The Institute of Industrial Applications Engineers, 2018.

[27]    L. Heflin, S. Walsh, M. Bagajewicz, Design of medical diagnostics products: A case-study of a saliva diagnostics kit, Comput. Chem. Eng. 33 (2009) 1067–1076.

[28]    H.E. Pople, Heuristic methods for imposing structure on ill-structured problems: The structuring of medical diagnostics, Artif. Intell. Med. 51 (1982) 119–190.

[29]    J.G. Richens, C.M. Lee, S. Johri, Improving the accuracy of medical diagnosis with causal machine learning, Nat. Commun. 11 (2020) 1–9.

[30]    N.A. Valentine, T.M. Alhawassi, G.W. Roberts, P.P. Vora, S.N. Stranks, M.P. Doogue, Detecting undiagnosed diabetes using glycated haemoglobin: an automated screening test in hospitalised patients, Med. J. Aust. 194 (2011) 160–164.

[31]    C. Jacobs, B. van Ginneken, Google's lung cancer AI: a promising tool that needs further validation, Nat. Rev. Clin. Oncol. 16 (2019) 532–533.

[32]    S. Kulkarni, N. Seneviratne, M.S. Baig, A.H.A. Khan, Artificial intelligence in medicine: where are we now?, Acad. Radiol. 27 (2020) 62–70.

[33]    A.Y. Sahib, H. Seyedarabi, R. Afrouzian, M. Farhoudi, A MATLAB-Based Toolbox to Simulate Transcranial Direct-Current Stimulation Using Flexible, Fast, and High Quality Tetrahedral Mesh Generation, IEEE Access. 10 (2022) 76573–76585.

[34] K. Doi, Current status and future potential of computer-aided diagnosis in medical imaging, Br. J. Radiol. 78 (2005) s3--s19.

[35] S.A. Naser, R. Al-Dahdooh, A. Mushtaha, M. El-Naffar, Knowledge management in ESMDA: expert system for medical diagnostic assistance, AIML J. 10 (2010) 31–40.

[36] K.-P. Adlassnig, A fuzzy logical model of computer-assisted medical diagnosis, Methods Inf. Med. 19 (1980) 141–148.

[37] F. Arute, K. Arya, R. Babbush, D. Bacon, J.C. Bardin, R. Barends, R. Biswas, S. Boixo, F.G.S.L. Brandao, D.A. Buell, others, Quantum supremacy using a programmable superconducting processor, Nature. 574 (2019) 505–510.

[38] P.T. Huynh, A.M. Jarolimek, S. Daye, The false-negative mammogram., Radiographics. 18 (1998) 1137–1154.

[39] W. Spiesberger, Mammogram inspection by computer, IEEE Trans. Biomed. Eng. (1979) 213–219.

[40] I.T. Gram, E. Lund, S.E. Slenker, Quality of life following a false positive mammogram, Br. J. Cancer. 62 (1990) 1018–1022.

[41] K. Doi, Computer-aided diagnosis in medical imaging: historical review, current status and future potential, Comput. Med. Imaging Graph. 31 (2007) 198–211.

[42] R.M. Nishikawa, Current status and future directions of computer-aided diagnosis in mammography, Comput. Med. Imaging Graph. 31 (2007) 224–235.

[43] K. Doi, H. MacMahon, S. Katsuragawa, R.M. Nishikawa, Y. Jiang, Computer-aided diagnosis in radiology: potential and pitfalls, Eur. J. Radiol. 31 (1999) 97–109.

[44] J.E. Meyer, D.B. Kopans, P.C. Stomper, K.K. Lindfors, Occult breast abnormalities: percutaneous preoperative needle localization., Radiology. 150 (1984) 335–337.

[45] K. Kerlikowske, R. Smith-Bindman, B.-M. Ljung, D. Grady, Evaluation of abnormal mammography results and palpable breast abnormalities, Ann. Intern. Med. 139 (2003) 274–284.

[46] A.E. McCart Reed, P. Kalita-De Croft, J.R. Kutasovic, J.M. Saunus, S.R. Lakhani, Recent advances in breast cancer research impacting clinical diagnostic practice, J. Pathol. 247 (2019) 552–562.

[47] R. Millstein, The logic theorist in LISP, Int. J. Comput. Math. 2 (1968) 111–122.

[48] C.W. Helstrom, Quantum detection and estimation theory, J. Stat. Phys. 1 (1969) 231–252.

[49] A.S. Holevo, Testing statistical hypotheses in quantum theory, Probab. Math. Stat. 3 (1982) 113–126.

[50] P. Benioff, Quantum mechanical Hamiltonian models of Turing machines, J. Stat. Phys. 29 (1982) 515–546.

[51] R.P. Feynman, others, Simulating physics with computers, Int. j. Theor. Phys. 21 (1982).

[52] R.P. Feynman, Quantum mechanical computers, Opt. News. 11 (1985) 11–20.

[53] A. Ekert, R. Jozsa, Quantum computation and Shor's factoring algorithm, Rev. Mod. Phys. 68 (1996) 733.

[54] A.Y. Kitaev, A. Shen, M.N. Vyalyi, M.N. Vyalyi, Classical and quantum computation, American Mathematical Soc., 2002.

[55] D.P. DiVincenzo, Quantum computation, Science (80-. ). 270 (1995) 255–261.

[56] F. Rosenblatt, The perceptron: a probabilistic model for information storage and organization in the brain., Psychol. Rev. 65 (1958) 386.

[57] H.-D. Block, The perceptron: A model for brain functioning. i, Rev. Mod. Phys. 34 (1962) 123.

[58] H.D. Block, B.W. Knight Jr, F. Rosenblatt, Analysis of a four-layer series-coupled perceptron. II, Rev. Mod. Phys. 34 (1962) 135.

[59] F. Rosenblatt, Perceptron simulation experiments, Proc. IRE. 48 (1960) 301–309.

[60] K.L. Kraft, Perceptron-like machines, University of British Columbia, 1969.

[61] B. Benaissa, N.A. Hocine, S. Khatir, M.K. Riahi, S. Mirjalili, YUKI Algorithm and POD-RBF for Elastostatic and dynamic crack identification, J. Comput. Sci. 55 (2021) 101451. https://doi.org/10.1016/j.jocs.2021.101451.

[62] S.-C.B. Lo, H.-P. Chan, J.-S. Lin, H. Li, M.T. Freedman, S.K. Mun, Artificial convolution neural network for

medical image pattern recognition, Neural Networks. 8 (1995) 1201–1214.

[63] M.-L. Huang, T.-Y. Lin, Dataset of breast mammography images with masses, Data Br. 31 (2020) 105928.

[64] N. Killoran, J. Izaac, N. Quesada, V. Bergholm, M. Amy, C. Weedbrook, Strawberry fields: A software platform for photonic quantum computing, Quantum. 3 (2019) 129.

[65] M. Broughton, G. Verdon, T. McCourt, A.J. Martinez, J.H. Yoo, S. V Isakov, P. Massey, M.Y. Niu, R. Halavati, E. Peters, others, Tensorflow quantum: A software framework for quantum machine learning, ArXiv Prepr. ArXiv2003.02989. (2020).

[66] P. Gokhale, J.M. Baker, C. Duckering, N.C. Brown, K.R. Brown, F.T. Chong, Asymptotic improvements to quantum circuits via qutrits, in: Proc. 46th Int. Symp. Comput. Archit., 2019: pp. 554–566.

[67] E. Farhi, H. Neven, Classification with quantum neural networks on near term processors, ArXiv Prepr. ArXiv1802.06002. (2018).

[68] C. Wang, A theory of generalization in learning machines with neural network applications, University of Pennsylvania, 1994.